\documentclass[conference]{IEEEtran}
\IEEEoverridecommandlockouts
\usepackage{cite}
\usepackage{amsmath,amssymb,amsfonts}
\usepackage{algorithmic}
\usepackage{graphicx}
\usepackage{textcomp}
\usepackage{xcolor}
\usepackage{listings}
\usepackage{verbatim}
\def\BibTeX{{\rm B\kern-.05em{\sc i\kern-.025em b}\kern-.08em
    T\kern-.1667em\lower.7ex\hbox{E}\kern-.125emX}}

\begin{document}

\title{Design of an embedded system with on-demand image capturing and transmission for remote agricultural monitoring
}

\author{Ivan Hu, Zihuai Lin\\
	School of Electrical and Information Engineering, The University of 
	Sydney, Australia\\
	Emails:	zihuai.lin@sydney.edu.au. 
}

\maketitle

\begin{abstract}
The ability to visually verify some element of a remotely controlled agricultural automation system through a photograph is valuable in many cases, not only in the operational phase of the system, but especially in the design and implementation phases. Owing to the remote location of many of the application sites, cellular techology is one enabling medium through which wireless transmission of the photographs could be realised. As data usage is a concern for systems using cellular technology, MQTT chosen  over other protocols due to its lower message-to-header overhead.  This paper outlines the hardware and firmware design of the LTE Cat-M1 enabled embedded system and the backend web development of the cloud-based web application to facilitate the receiving of the photograph. A satisfactory degree of implementation success was achieved in this project, with deployment to a production environment possible after further refinements. 
\end{abstract}
\smallskip

\section{Introduction}\label{sec:introduction}

Technological advancements constitute a crucial productivity multiplier in the agricultural sector, enabling expenditure reductions of time, manpower and material input for the same amount of yield.

Internet of Things (IoT) \cite{leng2020implementation,IoT_FD,RF_energy1} is a technology that can be leveraged to great advantage in the agricultural sector. This is particularly the case in Australia, where the vast geographical size of the continent coupled with low population density means that it is economically desirable to minimise human operators manually working on-site on farms, thereby avoiding overheads associated with travelling to, sometimes multiple, remote agricultural locations far away from the place of residence of the operator.

Through a network of sensors and actuators connected to a wireless network \cite{NC1,NC2,NC3,NC4,WRN,distributedRaptor,Raptor_ML,JNCC,RCRC,NC_book}, the operator has the power to remotely gather data such as precipitation, photosynthetically active radiation (PAR), and soil moisture about his/her farm. Using this information, the farmer could actuate nodes which could perform several functions, such as actuating a solenoid valve for irrigation/fertigation. Alternatively, actuators could be programmed to actuate automatically according to values reported by the sensor.

The degree to which the “smart farm” system \cite{digitalcattle} is productive depends greatly on the accurate and reliable operation of the sensors and actuators. For example, the valve actuators must accurately deliver a certain volume of water in order to effectively irrigate/fertigate a certain crop, but the only variable that the system has control over to achieve this goal is the duration of actuation. However, the relationship between duration of actuation and volume of water delivered cannot be generalised to all work sites because there are other variables at play which affects this relationship, such as the water pressure within the water input pipes. The degree to which this variable can vary depends on yet more factors, such as whether a water pressure regulation system is installed onsite \cite{waterpressure}. Thus, this relationship must be experimentally obtained in the implementation stage of the “smart system” for each farm.

Sensors must also work accurately and reliably because effective actuation depends on these measurements. For example, for the measure of precipitation, tipping-bucket rain gauges are known to exhibit significant errors if based on time scales of less than 10 to 15 minutes \cite{habib}. These issues
could be fixed to a satisfactory degree during the implementation stages of the system through the experimental processes of observation and calibration.

It would be greatly beneficial to the initial implementation process to have a method of visually verifying the correct operation of sensors and/or actuators in the “smart farm”. Engineers would implement the system according to theoretical specifications, and then remotely monitor the system for its actual behaviour before making the necessary adjustments in the next prototyping iteration, until the system behaves according to expectations.

An embedded device (hereafter known as the project transmission device) incorporating a camera, LTE-M Cat M1 modem and a SD card in conjunction with some software to receive the transmission, would be a solution to satisfy the above requirements.

The project transmission device would be electronically connected to the device being monitored using a digital I/O. After the monitored device performs its function, it would send a digital pulse signal to the project device, which would be powered on via a latching power switch circuit. The project device would then take a picture of the monitored subject, store it in the project device’s local SD card, perform encoding and encryption, upload the file to the cloud, and then power itself off by sending a digital pulse signal to the latching power switch circuit.

The outcome of this project would give engineers a method of remotely verifying the correct operation of the agricultural monitoring devices during the implementation stage without the hassle of going on-site. Not only that, the project device would also provide the ability to debug problems post-implementation, during the monitoring phase, if the “smart farm” system is observed to or suspected of exhibiting unexpected behaviour.

\section{Literature review\label{cha:litreivew}}

\subsection{LTE Cat-M1}
LTE-M, of which LTE Cat M1 is a subset, is a low power wide-area network (LPWAN) radio communications technology standard defined by the 3rd Generation Partnership Project (3GPP).

The design of LTE-M is motivated by a need to extend existing LTE technology \cite{cellular1,cellular2,cellular3,MIMO_capacity,network_capacity,UAVdownlink,UAV_THz,UAV_2} for better support of machine-type communications (MTC). As such, the specification is designed with some radio access principles in mind.

\subsubsection{Lower complexity and cost}

LTE-M was designed with the motivation to bring down costs of LTE modem production to a value comparable to GSM and GPRS, as at the time of conception, demands for MTC were adequately satisfied by those types of cheaper modems \cite{liberg2017cellular}.

3GPP Release 12 (Cat 0) was the first release of LTE-M and it has incorporated some elements of cost and complexity reduction as compared to Cat 1, an LTE technology. It is characterised by a reduced data rate of 1/1 (Mbps DL/Mbps UL) down from 10/5 (Mbps DL/Mbps UL), a single receive antenna down from two, and optional half-duplex frequency-division duplex (HD-FDD) operation \cite{liberg2017cellular}.

3GPP Release 13 (Cat M1), from the specifications of Cat 0, incorporates further cost and complexity reductions characterised as a reduced bandwidth of 1.4 MHz down from 20 MHz and an optional reduced transmit power of 20 dBm \cite{liberg2017cellular}.

\subsubsection{Coverage}

Nodes in IoT applications may be situated in locations where it is difficult to receive adequate coverage. Coupled with device simplification measures intended to drive down the complexity and cost of LTE-M modems, the device would experience significant coverage loss unless it is compensated with some coverage enhancement (CE) techniques.

CE techniques are used to increase the chance of successful transmission in low-coverage environments. In normal 4G operation, each transmission lasts approximately 1 millisecond. As such, a common coverage enhancement technique is the repetition of each transmission in order to increase the likelihood of successful transmission \cite{bergman_2017}.

LTE-M specifies two modes of CE, A and B, which respectively have increasing
coverage effects resulting from increasing degrees of repetition. 3GPP Release 13
mandates that Cat M1 devices must have support for CE mode A, with CE mode
B support being optional. CE mode B is always a superset of mode A.

In Australia, the Telstra network is able to provide LTE-M coverage to an extent that covers the overwhelming majority of Australia's arable land. 
\begin{figure}[htp]
\centering
\includegraphics[width=0.5\textwidth]{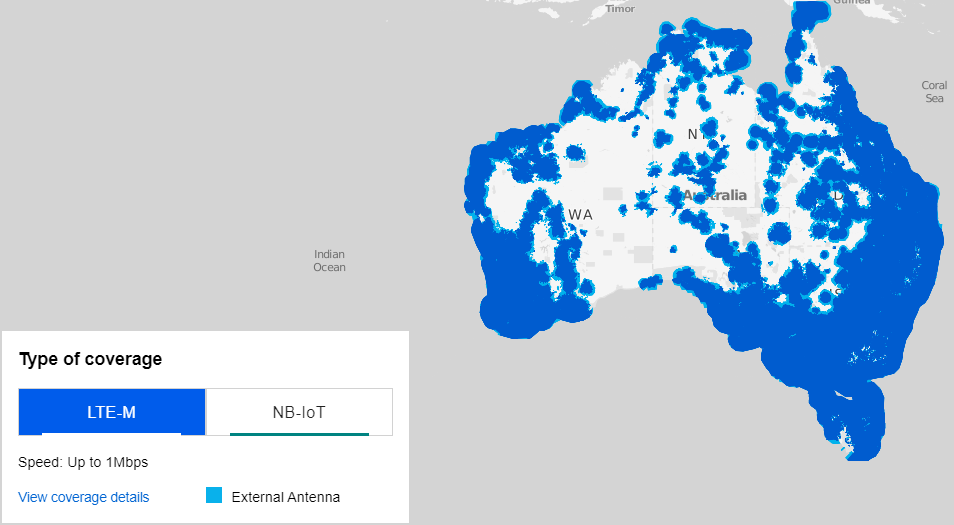}
\label{fig:figurecoverage}
\caption{Telstra network LTE-M coverage within Australia}
\end{figure}

\subsubsection{Low power usage}
Power Saving Mode (PSM) is specified for LTE-M in which the device enters a
state where it is unreachable by mobile terminated (MT) services, and as a result,
power consumption is reduced to a bare minimum. The LTE-M modem only leaves
PSM when a higher layer in the device triggers a mobile originated (MO) access,
for example for a data transmission \cite{liberg2017cellular}.

A further power optimisation feature of PSM is that in this mode, the device stays
registered to its network and preserves its connection configurations. This means
overheads associated with initial network attachment and connection setup are
avoided, as would be otherwise needed if the device were to experience a complete
power cycle \cite{liberg2017cellular}. Not only does this measure optimises the device for power consumption, but also lessens the time between initiation and completion of a MO
access.

After the data transmission associated with the MO access is complete, the device
enters a period of MT reachability as specified by its Active Timer. After the Active
Timer expires, the device once again enters the PSM state, remaining unreachable
by MT services until the next MO event, thus continuing the cycle \cite{liberg2017cellular}.

\subsubsection{Data rate}
According to \cite{liberg2017cellular} the estimated maximum uplink data rate for a Cat M1 device
supporting (half-duplex frequency division duplex) HD-FDD is 375kbps. Data rate
depends on the coupling loss experienced the device, with the maximum coupling
loss (MCL) data rate at 164 dB being 167 bps.

\subsection{MQTT}
Message Queuing Telemetry Transport (MQTT) is a protocol introduced in 1999,
having been developed by Andy Stanford-Clark and Arlen Nipper from IBM and
Arcom Control Systems Ltd respectively. The messaging protocol revolves around
a many-to-many relationship between clients performing publishing roles,
subscribing roles, or both, via a message broker. A publishing client publishes
messages to a broker, to addresses known as topics, from which subscribing clients
can subscribe to and receive messages from every topic \cite{naik2017}.

\subsubsection{Low messaging overhead}
MQTT, being designed for lightweight machine-to-machine (M2M)
communications in constrained networks \cite{bandy2013}, has a lower message to message
overhead ratio compared to other protocols such as AMQP and HTTP \cite{naik2017}.

Each MQTT control packet has a two-byte fixed header, with the 0th byte detailing
the type of control packet and flags specific to the control packet type, and the 1st
byte dedicated to specifying the remaining length of the packet. For every two-byte fixed headers sent, a payload of up to 256 MB can be supported.

\subsubsection{Quality of service (QoS) guarantee}
MQTT has three levels of QOS built into the protocol that offer three different
guarantees of message delivery.

QOS 0 guarantees that a message is received “at most once”. It is more intuitive
to think of this QOS level as an absence of guarantee rather than a guarantee,
because this is simply a fire-and-forget transmission. The recipient does not
acknowledge the reception of the message, nor does the sender store the message
so that it can be re-transmitted \cite{HiveMQqos}.

QOS 1 guarantees that a message is received “at least once”. As the name suggests,
there is a potential for the recipient to receive duplicate messages, and this isbecause the sender stores the message and continues to retransmit the message
until the sender receives an acknowledgement packet from the recipient \cite{HiveMQqos}. This
two-part handshake constitutes the middle ground out of the three QOS levels in
terms of overhead.

QOS 2 guarantees that a message is received “exactly once”. Utilising a more
complex and resource-expensive four-way handshake system than QOS 1, it
eliminates the issue of potential duplicate messages and guarantees that each
message is delivered exactly once. The sender first publishes the message to the
recipient and will continue to do so until a PUBREC message is received from the
recipient. PUBLISH packets subsequent to the initial one will have its duplicate
(DUP) flag set. The sender will eventually receive and store the PUBREC, at which
point the process is repeated with the final two handshakes PUBREL and
PUBCOMP. After this process, the receiver is left with one copy of the original
PUBLISH packet and the sender is assured that a copy of the original PUBLISH
packet has been sent \cite{HiveMQqos}.


\section{System overview}

\subsection{Typical operation}

The project transmission device lies in a powered-off state until it received a digital input pulse from the monitored device. The project transmission device will then power on and take a photograph. The photograph is stored in an SD card attached to the project transmission device as an JPEG file. The JPEG file is padded to the nearest file size that is divisible by 16 bytes. The JPEG file is encrypted with AES-128 using a pre-defined key. The encrypted file is encoded using Base64. The microcontroller sends a sequence of AT commands to the LTE-M modem to establish signal quality and establish a connection with an MQTT broker. The project transmission device will command the LTE modem to do a sequence of MQTT publishes which will publish the entire encrypted and encoded JPEG image in as many 1500-byte segments as required. 

On the receiver end, an active Python script (the "Downloader") running on the Linux backend of an AWS instance would have already established connection to the same MQTT broker and have subscribed to the same topic that the project transmission device publishes to. The Downloader downloads the encrypted and encoded image file onto the storage of the Linux server. Another python script (the "Decoder/Decrypter") is executed, which after the correct password is inputted, will decode the file from Base64 to an encrypted bitstream and then decrypt the file using the same pre-defined key. The Decoder/Decrypter writes original image onto the storage of the Linux server. The user interfaces with the receiver system via a webpage that is hosted on an Apache webserver that is running on the same Linux server. The webpage embeds two iframes containing respectively the Downloader and the Decoder/Decrypter, and an img element that displays the last image that has been downloaded, decoded and decrypted. The webpage serves as a rudimentary HTML-based dashboard.

\subsection{Development process}
This section gives a logically and roughly chronologically ordered outline of the development process. Each subsection will outline the justification and development process of a different technology used in this project. 

\subsubsection{Digitally-toggled power latch}
The use-case of the project transmission device is query-based, such that the system will come out of a dormant state based only on an external stimulus, as opposed to an internal wake-up source such as a timer. This means that the project transmission device need not have any computational capacity whatsoever during its dormant state. By exploiting this fact in conjunction with the nature of low duty-cycle embedded systems that most of their energy expenditure originates from dormant-state quiescent current,
significant power savings could be realised if a duty-cycling mechanism is chosen such that power is completely cut off from the project transmission device during its dormant state. 

The solution to this problem is the use of a latching power switch circuit that takes two digital signals as inputs to switch power on and off respectively. The circuit and its timing diagram is as follows.



The use of this circuit also has the additional benefit of saving development time, because there is no longer a need to configure the microcontroller or any peripheral components to go into their respective sleep modes via firmware. 

This circuit is adapted from \cite{eevblogslatch} with the adaption being the use of digital inputs instead of physical momentary switches as in the original circuit.

\subsubsection{Hardware development}
The AI-Thinker ESP32-CAM module is chosen as the main microcontroller module because it has an integrated camera, SD-card reader and an on-board 5V to 3.3V regulator. Furthermore, this module is officially supported by the Arduino libraries written for the ESP32 microcontroller, enabling ease of interfacing with the SD card and camera peripherals.

The Dragino NB-IoT Bee QG96 in conjunction with its NB-IoT Shield designed for the Arduino Uno is chosen as the cellular transmitter for the project transmission device. The Shield features on board 5V to 1.8V regulator which is useful because it can share the same 5V power rail with the ESP32-CAM module. Despite its name, the NB-IoT Bee features the Quectel BG96 modem which is primarily an LTE Cat-M1 modem. 

As the NB-IoT Shield is designed to interface with an Arduino Uno and not an ESP32-CAM, the two components had to be soldered onto a prototyping board with appropriate connections made using wires soldered to the underside of the board. The ESP32-CAM's UART port is connected to that of the Shield. Other than the 5V input power and ground rails, the only other notable connection is the 3.3V power rail between the two components as there is no on-board 3.3V voltage regulator on the Shield because it is designed to use the 3.3V power rail from its controlling board. 

The Shield and the ESP32-CAM are not directly soldered onto the board but are rather inserted into female header strips which are directly soldered onto the board. This modularity enhances the convenience of reprogramming the ESP32-CAM as the module must be inserted into a separate programming jig to do be reprogrammed. The programming jig features a generic FT232RL USB-to-TTL module and jumper wires to make the correct connections to the ESP32-CAM module.



\subsubsection{Debugging techniques}
JTAG-based debugging techniques commonly used in conventional embedded firmware development were not possible with the hardware used in the project transmission device. Although the ESP32 microcontroller possesses JTAG capabilities, the associated pins are already in use in the ESP32-CAM module to interface with the SD card and the camera \cite{espressifJTAG}. 

The relatively small number of pins (32) on the ESP32 family of microcontrollers means that the majority of pins in the ESP32-CAM modules are used to satisfy its I/O-hungry applications, namely to interface with the camera and the SD card. This means that there is only one UART port available to the user for development applications. For this application, that single UART port is used to interface with the LTE Cat-M1 modem for AT commands. This essentially eliminates the ability to debug through printing out variables. However, the ability to merely read AT commands between the microcontroller and the modem proved to have considerable debugging utility. This was achieved through the use of a TTL to USB converter by connecting the Rx of the TTL input to the Tx of the modem. Due to the fact that the modem echoes back any AT commands the microcontroller sends to it prior to sending back its response, merely monitoring this single connection is enough to obtain the information for the entire serial exchange between microcontroller and modem. The serial exchange can be monitored on a PC through any terminal software, which in this case was RealTerm. 




\subsubsection{Firmware development of cellular functionalities}
The source code is designed to be highly modular, with each module defined according to function. The three modules each govern the camera, cellular transmission and storage functions respectively. Each module has its own \verb|.cpp| and \verb|.h| file and its functions are called from \verb|ESP_mqtt_code.ino| in which the main loop of the program lies. The modular nature of the code aids in debugging during the development process because problems are isolated to that particular module. 


The firmware development process began with the development of the cellular functions module. The cellular module, like every other module in the code, has all its variables and buffers stored in an extern struct. This design feature is to emphasise the modular nature of the code. Being explicit during the process of reference reduces the potential for confusion. For example, if another module were to access the flags register of the cellular module, it would have to reference this variable using \verb|cellularObj.flags_reg|.


As the cellular module is essentially responsible for determining and sending the correct AT command based on the result of the previous AT command, the flow control of the cellular module is implemented as a finite state machine (FSM) with its state stored in an \verb|enum|. 


Execution of code during each state is done via a handler function which can repeatedly be called from the main loop. This has the additional benefit of the code being non-blocking. For example, if the handler function determines that the modem has not yet detected a signal, it can simply return from the function and try again the next time that the main loop calls the handler function.

Another design feature that is consistent with the non-blocking nature of this module is the use of non-blocking waits. Each AT command sent to the modem has a specific maximum response time that is specified by the Quectel BG96 application notes. For example, the \verb|AT+CSQ| command has a maximum response time of 300ms \cite{QuectelATmanual1,QuectelMQTTmanual2}. Therefore, the code will send the command in the \verb|cellular_send_CSQ| state and then store the number of milliseconds since beginning of code execution. In the next \verb|cellular_wait_CSQ| state, the code will check whether 300 milliseconds have elapsed since that time. Because the code is non-blocking, the main loop will keep calling \verb|cellularHandler()| for 300 milliseconds, after which point the FSM will progress into the \verb|cellular_do_CSQ| state where it will conduct flow control according to the response to the AT command, which the microcontroller should have received by now. Processing of all other AT commands are done in a similar manner.


A brief summary of the logical progression of the cellular module FSM is as follows. Start in a waiting state and wait for the modem to send a \verb|RDY| string. After that, send the \verb|AT+CSQ| check signal quality command to the modem and wait until the response is anything other than \verb|CSQ: 99,99| which should indicate that an acceptable signal has been acquired by the modem. The reason why the modem can directly look for a signal after powering on is because a one-time sequence of configuration AT commands has already been sent to the modem and the modem has stored the configuration settings into its non-volatile memory \cite{QuectelATmanual1,QuectelMQTTmanual2}


Then send \newline $\verb|AT+QMTOPEN=5,"broker.hivemq.com",1883|$ to connect to a public MQTT broker. A response other than \verb|QMTOPEN: 5,3| should indicate successful connection. Then send \verb|AT+QMTCONN=5,"clientExample"| to register a client connection to the same MQTT broker. This AT command will only be performed once, since the MQTT broker seems to react unfavourably to repeated \verb|AT+QMTCONN| connection requests. The firmware will instead query the broker for the connection status of the first connection request by using the \verb|AT+QMTCONN?| command, and a response of \verb|QMTCONN: 5,3| would indicate successful connection \cite{QuectelMQTTmanual2}

To perform a single MQTT publish on the topic \verb|testing|, the 
\verb|AT+QMTPUB=5,0,0,0,"testing"| command is sent. The modem will prompt the microcontroller to start sending the data payload by printing a \verb|>| character. After the microcontroller has sent its data payload, the microcontroller will signal end of data by sending a \verb|0x1A| (substitute character in the ASCII encoding) followed by a \verb|\r| (carriage return) \cite{QuectelMQTTmanual2}

The next sequence of AT commands deal with transmitting the file over a sequence of MQTT publishes. The microcontroller first publishes three comma separated values as a header. The header is necessary because the receiver must know this information in order to reconstitute the file properly. The first two values are the number of segments and the segment size respectively. The meaning of segment is the block of data that is to be published for the transmission of the actual file, because the entire file cannot be transmitted over a single MQTT publish. The modem can only support a maximum payload publish size of 1548 bytes \cite{QuectelMQTTmanual2}. In this firmware, the segment size is defined to be 1500. The number of segments is obtained by dividing the file size by 1500 and rounding up to the nearest integer. The third comma separated value is the size of the final segment, which can range from 1 to 1500.


After sending the header, the file is sent over a series of MQTT publishes each containing 1500 byte segment of the file with the exception of the last segment, which can contain 1 to 1500 bytes. After the file has been sent, the FSM will go into the \verb|cellular_done| state where it will send the \verb|AT+QLTS=2| command modem to obtain the date and time calculated from the latest time that the modem has obtained from the network \cite{QuectelATmanual1}. This last step is not strictly necessary but it was a development aid as it serves as an indicator that the cellular FSM has successfully entered the \verb|cellular_done| state. For further reference, Appendices A and B contains the full code for \verb|cellularFunctions.h| and \verb|cellularFunctions.cpp|.

\subsubsection{Python development of the file download functionality}
Chronologically speaking, toward the end of the cellular file transmission firmware development process, there was a need to develop the functionality for the receiver system in order to assure the functionality of the transmission system. For programming the receiver system, Python is chosen as a suitable language to expedite the development process, because it is widely and quantitatively regarded as a user-friendly language for basic procedural programs \cite{pythonfriendly} and because it has an abundance of libraries that is suitable for this application.  


$\verb|paho-mqtt|$ is a Python library that provides functionality for a program to act as an MQTT client. 
The code structure consists of a run-once section prior to \newline \verb|client.loop_forever()| which serves as a infinite loop to keep the program open.

\verb|on_connect()| and \verb|on_message()| are functions which practically serve as Python equivalents to C interrupt routines. \verb|on_connect()| is called when the program has successfully connected to an MQTT broker. In this case, a message is printed out signifying the success/failure of the connection and the client promptly subscribes to the same topic that the transmitter would be publishing the file on.

The main bulk of the development involves writing the \verb|on_message()| function which handles parsing the series of MQTT publishes that make up the header and the segmented payload of the transmission. There was a need for \verb|on_message()| to access variables persistent to numerous different function calls during this process. The dictionary \verb|client_userdata| serves as a single memory location where different variables could be accessed. \verb|client_userdata| is called with the \verb|global| keyword within \verb|on_message()| so that the function could modify the variable which is located outside of the function scope. 

The first part of the function handles parsing of the header. The payload of the MQTT publish containing the header is stripped of trailing characters so it becomes a true comma-separated string of 3 integers containing the number of segments, segment size and last segment size respectively. This comma-separated string is stored within \verb|client_userdata|.

The next part of the function handles the case where the incoming MQTT publish is the first segment of the data payload. The function will open a file in binary writing mode and write the first 1500 bytes of the file. \verb|currentSegment| within \verb|client_userdata| is incremented to keep track of how many payload segments the program has received thus far. The next case handles an MQTT data payload publish that is neither the first nor last segment. In this case, the program will open the same file in binary append mode and append the next 1500 bytes to the file. The last case handles an MQTT data payload publish that is the last segment. In this case, the program will open the same file in binary append mode and append the last of the bytes into the file, and then resets \verb|client_userdata| to its initial state so that the program will be ready for a subsequent file download sequence. 

\subsubsection{Firmware development of file processing functionalities}
To address data security concerns, files should always be encrypted prior to transmission over the internet. 




It was discovered that raw binary data, encrypted or not, cannot be fed directly into the LTE modem and published using MQTT. Doing so results in unsuccessful publishes. The reason for this is not definitively proven but it is likely due to the fact that some binary sequences can be interpreted by some intermediary network protocol to be special characters, and as a result breaking the protocol.   
The \verb|crypto/base64.h| 
library encrypts buffers containing binary data of any size into another buffer of greater size than the original due to the nature of the encoding. It will even return the size of the output buffer to aid the user's development process. The  \verb|void base64Encode(fs::FS &fs, char* srcDir,| \verb| char* destDir)| function was written to encode a binary file into a file that is wholly represented by ASCII characters. The operation of the function is similar to \verb|aesEncode()|, except that the user-padding requirement is not a concern here because this library handles padding. The RFC 4648 standard of Base64 encoding specifies that padding will be added if at the end of the data being encoded, fewer than 24 bits are available \cite{base64standard}. Appendices E and F contain the full code for this module.

\subsubsection{Python development of the file decoding and decryption functionalities}
A Python program was subsequently developed to provide a means of recovering the original file from its encrypted and encoded analogue. 


As a proof of concept of data privacy, any users of this program must first input the correct password. The severe security limitations of this implementation of password authentication will be addressed in the next chapter. 

The \verb|base64| package within the Python Standard Library provides functionality for Base64 decoding. As Python is a high-level language that is being run on a processor with significantly higher processing capabilities than an ESP32, the entire \verb|inputjpg_encrypted_encoded| file can be opened and its entire content can be stored in a single variable. The package then takes this variable and outputs the the decoded binary content into another variable. The program then writes this binary stream into a file called \verb|image_encrypted|.

The \verb|PyCryptodome| package provides various cryptographical tools for the Python language. The subpackage \verb|Crypto.Cipher| is used for AES128 decryption. As is mentioned earlier, 
the Base64 encoding process could potentially insert padding that will cause the file size to no longer be divisible by 16. Due to this, the program calculates the next lower number from the current file size that is divisible by 16. The program will then truncate the file to this size, then input it into the \verb|Crypto.Cipher| library. The result is the recovered JPEG image which to human eyes, has no discernible difference to the original image that is taken by the embedded system.  


\subsubsection{Development of a HTML-based dashboard}
To enhance the device compatibility of the system e.g. if the user wanted to download, decode and decrypt transmission from a device which cannot natively run Python, these functions must be integrated into a web-based platform. 

Due to time constraints, it was determined that the most time-effective solution to this requirement was to somehow run the Python code already developed from a web page. An AWS Linux instance was set up to host the website through Apache. On the webpage, two HTML iframes embedding two SSH terminals to the Linux instance are implemented using  \verb|shellinabox|. 
This is achievable because \verb|shellinabox| is a webserver that implements and emulates an SSH terminal connection on port 4200. In this manner, the user can, from a single webpage, run the two aforementioned Python scripts for downloading and decoding/decrypting respectively.

Additionally, from the same webpage, the user can view the recovered image. A "Refresh Image" button allows the user to refresh the image for subsequent downloads without refreshing the whole page so as not to compromise the login status of the two \verb|shellinabox| iframes. The "Refresh Image" button calls a simple Javascript cache breaker function that retrieves the newly downloaded image and updates the \verb|<img>| element. Without this cache breaker function, the user's web browser would simply display the same image that is stored in the web browser's cache. 


\subsubsection{Firmware incorporation of the camera function}
The most trivial part of the development is left until last. There already exist numerous tutorials and code examples on the internet regarding how to take a picture on an ESP32-CAM and store the result in the SD card. The code in this project is adapted from \cite{randomnerdcamera}. The \verb|esp_camera.h| library 
forms the main basis of this code.

With regards to the code, it is self-explanatory and there is not much that requires elaboration. See Appendices I and J for the full code.

\section{Results and Discussion\label{cha:results}}

\subsection{Cellular data usage}
Using the file transmission scheme developed in this project, a 18093 byte file is able to be transmitted over the internet with the cellular service provider claiming the total volume of data transmitted to be 20686 bytes.



The transmission scheme designed in this project is able to achieve a 87.46\% payload to total data usage ratio. From inspection, this figure looks quite optimised. To fully appreciate the efficiency gains of using MQTT as opposed to HTTP, it would be ideal to replicate this project using HTTP, and compare the results. However, this is outside the scope of this project. 

\subsection{Image characteristics}
The camera is configured to capture a picture 640 by 480 pixel in dimension. The image quality in terms of lossiness is set to 10 from a scale of 1 to 63, with a lower number indicating higher quality. 
The image quality is good enough to discern black, size 72 Arial font printed on white paper from a distance of 1.5 meters. 

\begin{figure}[htp]
\centering
\includegraphics[width=0.45\textwidth]{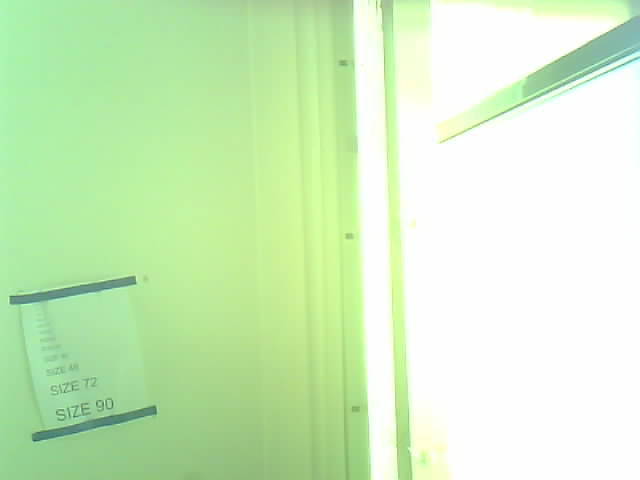}
\label{fig:figure11}
\caption{The captured image.}
\end{figure}

Image resolution and quality is a compromise between visual information and cellular data usage. Depending on the specific application for the project transmission device, the \verb|esp_camera.h| library allows for easy configuration of these parameters. 

\subsubsection{End-to-end process duration}
To transmit a 18093 byte file, the project transmission device takes approximately 40 seconds from the time it is switched on to when the entire file is received on a Sydney-based AWS server \cite{projectdemovideo}. This duration is subject to variation depending on the size of the image captured, the wireless channel between the project transmission device and its base station, and the state of the IP route that the datagrams will traverse over the internet.

The following is an analysis on the factors contributing to the end-to-end process duration. 

There are 26 seconds between when the project transmission device is turned on and when it starts uploading. For the most part, these 26 seconds cannot be further optimised, because of, firstly, factors out of my control and secondly, factors that have been optimised to the fullest extent without sacrificing system reliability.  

Of these 26 seconds, 5 seconds each are allocated to waiting for the responses of the \verb|AT+QMTOPEN| and \verb|AT+QMTCONN| commands respectively. These allocated times are chosen through repeated experimentation to find the best compromise between waiting too long or not waiting long enough for the response to be intercepted. Such experimentation is necessary for these commands because the LTE modem's user manual states that the Maximum Response Time (MRT) for these commands are 75 seconds \cite{QuectelMQTTmanual2}, and it would certainly be infeasible to wait the entire duration of the MRT for these commands. Another factor that has already been optimised is the 3 seconds waiting time for the \verb|RDY| string from the modem.

However, for some other AT commands, for instance \verb|AT+CSQ| have a MRT of 300ms. In this case, it would be feasible and best practice to wait the entire duration. The same applies for all AT commands with a specified MRT in the order of less than a few seconds. 

The final factor in this 26 second duration that is entire out of my control is waiting for \verb|AT+CSQ| to return a signal strength signifying connection to the base station. This is entirely dependant on the wireless channel quality between the project transmission device and the base station. 

For the remainder approximately 14 seconds, the project transmission device repeatedly publishes 1500 byte segments of the file. This time can be divided into actual transmission time which is dependant on the baud rate of the serial port between the microncontroller and the LTE modem, and the waiting time for the response of the moodem to the \verb|AT+QMTPUB| command. Since the baud rate is 115200 with 8 data bits and one stop bit, the true data rate is 102400 bits per second. Therefore, it takes approximately 1414 milliseconds to transmit 18093 bytes of data plus a 10 byte header. And since the header plus payload is divided into 14 publishes in total with a 500ms wait time after each publish, 7000 milliseconds is spent during this waiting period. It is assumed that the rest of the unaccounted-for time within this 14 second duration are due to other latency factors out of my control.  

Therefore, it can be concluded that the main factor (7 seconds out of 14) contributing to the time spent during the transmission phase is the AT command response waiting time. As the user guide specifies an unrealistically long MRT of 15 seconds for \verb|AT+QMTPUB| \cite{QuectelMQTTmanual2}, it has been experimentally found that the lowest safe waiting time is 500ms and it would be unwise to compromise system reliability by reducing it further. Naturally, as the size of the file being transmitted increases, the number of separate MQTT publishes to transmit the entire file increases, leading to the proportion of time taken by the AT command response waiting time also increasing. This means that optimisation of serial transmission time through increasing baud rate becomes less impactful as the file size increases. Assuming that the image resolution and quality (and therefore size) should not be further reduced in order to preserve visual information, it can be concluded that little can be done to optimise transmission time. 

\subsection{Power usage}
When the project transmission device is turned on, its average current consumption from the power source is approximately $190$ milliampere (mA). 

Due to time constraints, the power latch circuit has not been successfully implemented. However, simulation of the power latch circuit shows that it consumes a quiescent current of $8.885$ microampere ($\mu
$A) (see Appendix L).

Based on these figures, if the project transmission device is to perform its functions three times every hour, with the remainder of the time being turned off, the project transmission device would have an average current draw of:

\[ \frac{3\times0.19\times40+8.885\times10^{-6}\times3480}{3600}\approx6.3 mA \]

Accounting for the 5V power source, the project transmission device draws an average of $0.0315$ {watt}.

To calculate the battery life if a project transmission device were to use a typical 18650 lithium-ion battery and a boost converter to 5V, let us consider the following parameters. Suppose the battery has a current capacity of $2600$ milliampere hour (mAh) at 3.7V, and the MCP16251 boost converter with a typical efficiency of 96\% \cite{boostconverter} is used. The battery capacity would have a power capacity of 9.62Wh. This means it can sustain the project transmission device for over 290 hours. If we consider solar energy as a means of energy harvesting for this application, it would take over 12 days of zero solar input to deplete the battery. 

Therefore, the energy consumption of this device is low enough such that it is feasible to be deployed at a remote location with solar energy harvesting to enable continuous operation from an energy consumption standpoint. 

\subsubsection{Data security}
Data security \cite{security1,security2} in the form of end-to-end encryption is crucial in this project because the project transmission device, by default, is configured in firmware to use a public MQTT test broker \verb|broker.hivemq.com| on a public test topic \verb|testing| with no MQTT-layer authentication. This means that anyone connected to this broker can subscribe to the same topic and be able to have access to any data that the project transmission device publishes. It is therefore crucial from a data security standpoint that the project transmission device encrypts any published data so that even if the encrypted data is intercepted, no information can be discerned from it. It is not sufficient to rely on Base64 alone to obfuscate the data because this encoding scheme is ubiquitously used and can be recognised easily by inspection.

The fact that the MQTT broker acts as an intermediary between the project transmission device and the receiver enhances the data security of the system as a whole. In a traditional source-to-destination data transmission protocol, for instance HTTP using a POST request, data interception would result in information pertaining to the address of the destination to be divulged to the attacker. Once this is known, the attacker can then attempt to find and exploit vulnerabilities at the destination. However, to attack this project's system, the most likely manner that malicious attackers would attempt to intercept data is by monitoring inbound traffic at the broker. But after this, it becomes very difficult for the attacker to work out the address of the destination, because the the data is directed to the rightful receiver of the data by way of a client-oriented MQTT subscription. The only way that the attacker could conceivably work out the destination address is by sampling all outbound traffic from the MQTT broker and analyse it all, which would be a very impractical task in the case of a public MQTT broker. 

Changing topics from MQTT to the web-based dashboard at the receiver, the implementation of the dashboard presents a very serious security vulnerability. In order for a user to be able to use the downloader and Decoder/Decrypter functions, the user must be given root access to the Linux backend of the receiver system. Should the password-protected root access be compromised, the attacker can directly access the Python scripts which contain the AES128 key as an unobfuscated string. To address this issue, the Python scripts providing the downloader and Decoder/Decrypter functionalities should be ported to a proper web framework and any notion of being able to SSH into the backend of the server should absolutely be removed from the front end i.e. the webpage. 

\subsubsection{Data privacy}
The concept of data privacy \cite{privacy1,privacy2,privacy3,privacy4} is addressed by this project on a proof-of-concept level through the separation of the download versus decoding/decryption functions at the receiver. Having a separate password before access to the decoding/decryption functions serves as a representation that there should be two levels of security clearance for the personnel manning the receiver system; one to download and one to actually recover the originally-sent data. 

The implementation used in this project is quite vulnerable. The password is simply stored as an unobfuscated string within the Python script for the Decoder/Decrypter which can easily be obtained by opening the script with a text editor. A more robust password authentication system is required for proper implementation of data privacy measures. 


\section{Conclusion\label{cha:conclusion}}

This project has successfully implemented an embedded system that is able to, upon powering on, take a photograph, encrypt and encode the photograph, publish the photograph as a series of MQTT publishes to an MQTT broker. This project has also successfully implemented a web application that is able to subscribe to and "listen for" the file at the same MQTT broker, download the file, and then perform decoding and decryption in order to recover the original photograph. 

The use of MQTT as a protocol in the context of file transmission is unorthodox, but theoretically provides data usage efficiency gains compared to other protocols such as HTTP. It also provides a slight edge in data security over HTTP as incoming MQTT backets to the broker does not store any information about the address of the intended recipient.

The use of a digital pulse-trigger latch circuit as a means of switching power on and off to the project device was not realised, but simulations show that it is a functional solution. 

With no fundamental conceptual shortcomings which could not be fixed with further refinements, the development undertaken within the course of this project can be eventually used in a real-life production environment.

\end{document}